\documentstyle[epsfig]{aa}

\def\beq{\begin{equation}}
\def\enq{\end{equation}}
\def\bea{\begin{eqnarray}}
\def\ena{\end{eqnarray}}
\def\bec{\begin{center}}
\def\enc{\end{center}}

\def\ergcm2si{\hbox{ergs~cm$^{-2}$s$^{-1}$}}

\begin{document}

\title{XMM-Newton studies of a massive cluster of galaxies: 
\\
RXCJ2228.6+2036}

\author{S.M.~Jia\inst{1,2}, H.~B\"{o}hringer\inst{1}, E.~Pointecouteau\inst{3},
Y.~Chen\inst{2}, and Y.Y.~Zhang\inst{4}}

%\offprints{S.M.~Jia (E-mail: jiasm@mail.ihep.ac.cn)}

\institute{
Max-Planck-Institut f\"{u}r extraterrestrische Physik, 85748 Garching, Germany
\\
e-mail: jiasm@ihep.ac.cn
\and
Key Laboratory of Particle Astrophysics, Institute of High Energy Physics, 
Chinese Academy of Sciences, Beijing 100049, P.R. China
\and
CESR-CNRS, 9 Av. du Colonel Roche, 31028 Toulouse, France 
\and
Argelander-Institut f\"{u}r Astronomie, Rheinische Friedrich-Wilhelms-Universit\"{a}t 
Bonn, Auf dem Huegel 71, D-53121 Bonn, Germany
}

\date{Received  / accepted }

\abstract{ We present the X-ray properties of a massive cluster of galaxies 
(RXCJ2228.6+2036 at $z=0.421$) using {\it XMM-Newton} data. The X-ray mass 
modeling is based on the temperature and density distributions of the 
intracluster medium derived using a deprojection method. We found that 
RXCJ2228.6+2036 is a hot cluster ($T_{500}=8.92^{+1.78}_{-1.32}$ keV) showing 
a cooling flow rate of $12.0^{+56.0}_{-12.0}$ M$_{\odot}$yr$^{-1}$ based on
spectral fitting within the cooling flow radius ($r_{cool}=147\pm10$ kpc). The 
total cluster mass is $M_{500}=(1.19\pm0.35)\times10^{15}$ M$_{\odot}$ and the 
mean gas mass fraction is $f_{gas}=0.165\pm0.045$ at $r_{500}=1.61\pm0.16$ Mpc. 
We discussed the PSF-correction effect on the spectral analysis and found that 
for the annular width we chose the PSF-corrected temperatures are consistent 
with those without PSF-correction. We observed a remarkable agreement between 
X-ray and SZ results, which is of prime importance for the future SZ survey. 
RXCJ2228.6+2036 obeys the empirical scaling relations found in general massive 
galaxy clusters (e.g. $S$--$T$, $M$--$T$, $L$--$T$ and $M$--$Y$) after 
accounting for self-similar evolution.

}

\titlerunning{XMM-Newton studies of a massive cluster of galaxies: 
RXCJ2228.6+2036} \authorrunning{S.M. Jia et al.}

\maketitle

%%%%%%%%%%%%%%%%%%%%%%%%%%%%%%%%%%%%%%%%%%%%%%%%%%%%%%%%%%%%%%%%%%%%%%%%%%%%%%%
\section{Introduction}
The gravitational growth of fluctuations in the matter density distribution can
be traced by the evolution of the galaxy cluster mass function (e.g. Schuecker
et al. 2003). The hot and distant clusters are at the upper end of the mass 
distribution, thus they can be used to probe the cosmic evolution of 
large-scale structure and are therefore fundamental probes for cosmology. But 
to date still very few hot and distant clusters are known. Therefore, it is 
important to study such clusters in detail, especially in X-ray.

RXCJ2228.6+2036 is one of the distant ({\it z} = 0.421) and X-ray luminous 
clusters of galaxies in the northern sky. It is suspected to be massive and hot,
and was well recognized as an extended X-ray source in the ROSAT All-Sky Survey, 
included in both the NORAS galaxy cluster survey (B\"{o}hringer et al. 2000) 
and the ROSAT Brightest Cluster Sample (Ebeling et al. 2000).

The first combined SZ versus X-ray analysis for RXCJ2228.5+2036 is based on the
SZ data from the Nobeyama Radio Observatory (NRO) 45 m radio telescope and the
X-ray data from ROSAT/HRI. It shows that RXCJ2228.6+2036 is a hot and massive
cluster with $T = 10.4 \pm 1.8$ keV, $M_{tot}(r<R_v=r_{178}=2.9$ Mpc) =
$(1.8\pm0.4)\times10^{15}$ M$_{\odot}$, and with a gas mass fraction of $f_{gas}
= 0.22 \pm 0.06$ (Pointecouteau et al.  2002). Recently, LaRoque et al.
(2006) performed the {\it Chandra} X-ray versus OVRO/BIMA interferometric SZ
effect measurements for the same cluster, giving $T = 8.43^{+0.78}_{-0.71}$
keV, $f_{gas} = 0.138 \pm 0.009$ from the X-ray data, and $f_{gas} =
0.188^{+0.035}_{-0.031}$ from the SZ data at $r_{2500}$. RXCJ2228.6+2036, as one
of the clusters in the sample of X-ray luminous galaxy clusters with both X-ray
({\it Chandra}) and SZ observations in Morandi et al. (2007), has a
temperature of $T = 6.86^{+0.89}_{-0.71}$ keV and a total mass of
$M_{tot}=(4.90\pm4.35)\times10^{14}$ M$_{\odot}$ at $r_{500}=1033 \pm 464$
kpc. However, the above results are all based on the mass modeling under the
assumption of isothermality of the ICM. The {\it XMM-Newton} EPIC instruments 
have both high spatial and spectral resolutions and a large field of view, and 
are therefore suitable for a spatially resolved spectral analysis. We make use
of {\it XMM-Newton} observations to carry on a detailed study of RXCJ2228.6+2036 
based on the X-ray mass modeling using a spatially resolved radial temperature 
distribution and perform a detailed X-ray versus SZ comparison.

The structure of this paper is as follows: Sect. 2 describes the data,
background subtraction method and spectral deprojection technique. Sect. 3
presents the spectral measurements using different models to derive the radial
temperature profile, cooling time and mass deposition rate. In Sect. 4 we show
the radial electron density profile and X-ray mass modeling. In Sect. 5 we
discuss the impact of the PSF correction on the spectral analysis, and compare
RXCJ2228.6+2036 to the SZ measurements and the empirical scaling relations for 
massive galaxy clusters. We draw our conclusion in Sect. 6.

Throughout this paper, unless explicitly stated otherwise, we use the 0.5-10
keV energy band in our spectral analysis. The cosmological model used is 
$H_0$ = $70$ km s$^{-1}${Mpc}$^{-1}$, $\Omega_m$ = 0.3, and $\Omega_{\Lambda}$ 
= 0.7, in which 1$^{\prime}$ corresponds to 332.7 kpc at the distance of 
RXCJ2228.6+2036.

%%%%%%%%%%%%%%%%%%%%%%%%%%%%%%%%%%%%%%%%%%%%%%%%%%%%%%%%%%%%%%%%%%%%%%%%%%%%%%%
\section{Observation and data reduction}
\subsection{Data preparation}

RXCJ2228.6+2036 has been observed for 26 ksec in November 2003 by 
{\it XMM-Newton} and its observation ID is 0147890101. For our purpose, we only
use EPIC data (MOS1, MOS2 and pn). The observations are performed with a thin
filter and in the extended full frame mode for pn and the full frame mode for
MOS. Throughout this analysis, we only use the events with FLAG=0, and with 
PATTERN$\le$4 for pn and PATTERN$\le$12 for MOS. The reduction was performed 
in SAS 7.1.0.

The light curve of the observation shows some flares (i) in the hard band
(above 10 keV for MOS and above 12 keV for pn), possibly caused by the
particle background, and (ii) in the soft band (0.3-10 keV), possibly due to
the episodes of `soft proton flares' (De Luca \& Molendi 2004). Therefore both
the hard and the soft bands are used to select the good time intervals
(GTI) as described in Zhang et al. (2006). The GTI screening procedure gives us
22 ks MOS1 data, 22 ks MOS2 data, and 18 ks pn data.

We applied the SAS task {\it edetect\underline{~}chain} to detect the point 
sources {\bf (the radius of the point sources is $0.6'$ containing 93\% flux
from the point source)}, and excised all of them from the cluster region. 
Then, a SAS command {\it evigweight} was used to create the vignetting weighted 
column in the event list to account for the vignetting correction for the 
effective area.

Due to read-out time delay, the pn data require a correction for the
Out-of-time (OOT) events. We created the simulated OOT event file and used it 
to account for this (see Str\"{u}der et al.  2001 ) in our analysis.

\subsection{Background subtraction}
The {\it XMM-Newton} background normally consists of the following two 
components. i) Particle background: high energy particles such as cosmic-rays 
pass through the satellite and deposit a fraction of their energy on the 
detector. It dominates at high energies and shows no vignetting. ii) Cosmic 
X-ray Background (CXB): the CXB varies across the sky (Snowden et al. 1997). 
It is more important at low energies and shows a vignetting effect.

We choose the blank sky accumulations in the Chandra Deep Field South (CDFS)
as the background (Zhang et al. 2007), which was also observed with the thin
filter. We applied the same reduction procedure to the CDFS as to the cluster
in the same detector coordinates, and the effective exposure time we obtained
for CDFS is 54 ks for pn, 61 ks for MOS1 and 61 ks for MOS2. RXCJ2228.6+2036 
is a distant cluster ($z=0.421$), {\bf and we estimated that the signal-to-noise 
ratio of the region $6'<R<6.5'$ is about 20\%, so we can approximately assume 
that the emission of the cluster only covers the inner part of the field of 
view ($R<6'$)}. The outer region ($6.5'<R<8'$) can thus be used to monitor the 
residual background. Here we applied a double-background subtraction method to 
correct for these two kinds of background components as used in Arnaud et al. 
(2002). First we estimate the ratio of the particle background, $\alpha$, 
between RXCJ2228.6+2036 and CDFS from the total count rate in the high energy 
band (10-12 keV for MOS and 12-14 keV for pn), as described in Pointecouteau et 
al.  (2004). $S_0$ and $B_0$ are the background spectra of the cluster and CDFS 
in the region of $6.5'<r<8'$ with an area of $A_0$, and $S_i$ and $B_i$ for 
spectra in the $i$th ring of the cluster and CDFS with an area of $A_i$. Then 
the cluster spectrum, after the double-background subtraction, $S(i)$ is (e.g. 
Jia et al.  2006, Zhang et al. 2006):
\begin{equation}
S(i)=S_i-\alpha B_i-\frac{A_i}{A_0}(S_0-\alpha B_0)
\end{equation}

\subsection{Spectral deprojection}
The combined image of MOS1 and MOS2 in the energy band 0.5-10 keV is shown in
Fig. 1. It is corrected for vignetting and smoothed with a maximum Gaussian
smoothing size of $\sigma = 5$ pixels. As shown in this figure, the X-ray 
emission of RXCJ2228.6+2036 appears to be extended and almost symmetric except 
for some bright point sources (which were subtracted before the spectrum 
extraction, see Sect. 2.1). We extract the spectra from annular regions 
centered on the emission peak, and the width of each ring is determined 
according to the criterion described in Zhang et al. (2007): $\sim$2000 net 
counts per bin in 2-7 keV to get a temperature with $\sim$15\% uncertainty. 
Considering the PSF (Point Spread Function) effect of {\it XMM-Newton} EPIC, 
whose Full Width at Half Maximum (FWHM) is $5''$ for MOS and 6$\arcsec$ for pn, 
the minimum width of each ring was set at $0.5'$. We thus obtained 5 annuli to 
extract spectra out to $6'$.

\begin{figure}
\centerline{\psfig{file=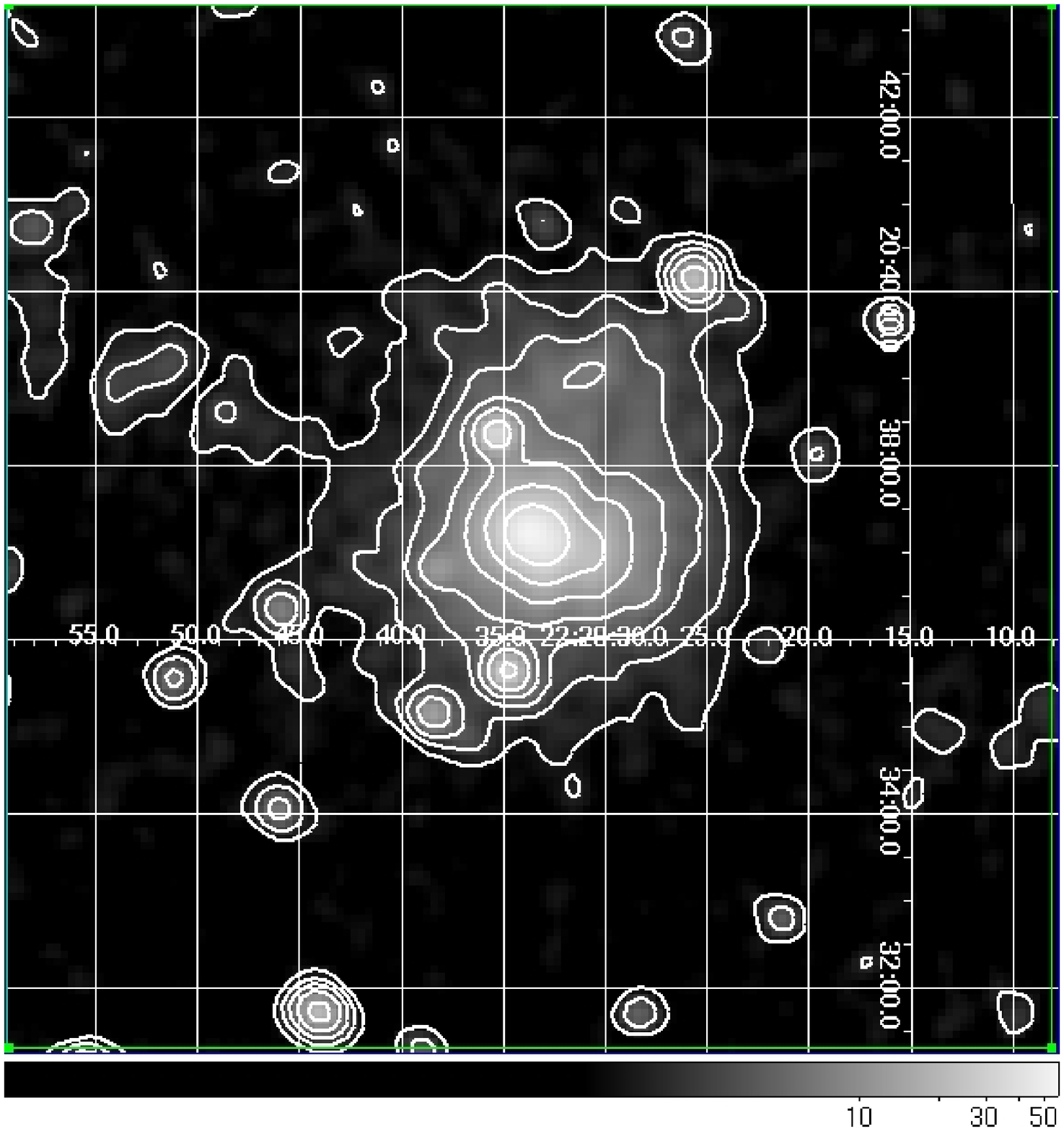,width=8cm}}
\caption{The combined image of MOS1 and MOS2 of RXCJ2228.6+2036 ($12'\times12'$) 
corrected for vignetting and smoothed with a maximum Gaussian smoothing size 
of $\sigma = 5$ pixels.}
\end{figure}

The deprojected spectra are calculated by subtracting all the contributions
from the outer regions. Within each radial range, we assume the same spectrum
per unit volume. The deprojected spectrum of the $i$th shell is then
calculated by subtracting the contributions from the $i$+1th shell to the
outmost one (e.g. Matsushita et al. 2002 and Nulsen \& B\"{o}hringer 1995).
The detailed calculation procedures are described as in Jia et al. (2004,
2006):
\begin{equation}
D(i)=\left[S(i)-\sum_{k=i+1}^{N}C_v(k,i) \cdot D(k)\right]/C_v(i,i),
\end{equation}
here $D(i)$ is the deprojected spectrum of the $i$th shell, $S(i)$ is the
double-background subtracted spectrum of the $i$th shell and $C_v(k,i)$ is 
the fraction of the volume of the $k$th shell projected to the $i$th ring.

The on-axis rmf (response matrix file) and arf (auxiliary responds file) are
generated by the SAS task {\it rmfgen} and {\it arfgen} and are used to 
recover the correct spectral shape and normalization of the cluster emission 
components.
%We generate both the response matrix file (rmf) and auxiliary responds 
%file (arf) by the SAS task {\it rmfgen} and {\it arfgen} and use them to 
%recover the correct spectral shape and normalization of the cluster emission
%components.

%%%%%%%%%%%%%%%%%%%%%%%%%%%%%%%%%%%%%%%%%%%%%%%%%%%%%%%%%%%%%%%%%%%%%%%%%%%%%%%
\section{Spectral analysis}
\subsection{Radial deprojected temperature profile}
The spectral analysis is carried out in XSPEC version 11.3.2 (Arnaud 1996).
To study the temperature distribution of RXCJ2228.6+2036, we perform a joint fit
to the spectra of pn and MOS with an absorbed Mekal model:
\begin{equation}
Model_1=Wabs(N_H)\times Mekal(T,z,A,norm),
\end{equation}
in which Wabs is a photoelectric absorption model (Morrison \& McCammon 1983)
and Mekal is a single-temperature plasma emission model (Mewe et al.1985,
1986; Kaastra 1992; Liedahl et al. 1995). The temperature $T$, metallicity $A$
and normalization (emission measure) $norm$ are free parameters. We fixed the
redshift $z$ to 0.421 and the absorption $N_H$ to the Galactic value
4.68$\times$10$^{20}$ cm$^{-2}$ (Dickey \& Lockman 1990).  The fitting results
are listed in Table 1 and the central spectra fitted by this model are shown
as Fig. 3(a).

\begin{table*}
\caption{The best-fit free parameters of RXCJ2228.6+2036 by the
single-temperature model: the temperature $T$; the abundance $A$ and the 
normalized constant $norm$ for the simultaneously fitting of pn and MOS. 
$norm = 10^{-14}/(4\pi (D_A\times(1+z))^2) \int n_e n_H dV$, where $D_A$ is 
the angular size distance to the source (cm) and $n_e$ is the electron 
density (cm$^{-3}$). $L_{bol}$ represents the bolometric luminosity (0.01-60
keV) in the units of $10^{44}$ erg s$^{-1}$. The errors represent a confidence 
level of 90\%.}
\begin{center}
{\footnotesize
\begin{tabular}{cccccc}
\hline
 $r$($'$) & $T$ (keV) & $A$ (solar)& $norm$($10^{-3}$cm$^{-5}$)
& $\chi^{2}_{\mathrm{red}}$/$dof$ & $L_{bol}$($10^{44}$ erg s$^{-1}$)
\\
\hline
0.0-0.5 & $8.26^{+1.02}_{-0.92}$ & $0.31^{+0.19}_{-0.18}$ & $0.72\pm0.03$ & 
1.12/251 & $4.65^{+0.38}_{-0.31}$
\\
0.5-1.0 & $11.65^{+2.54}_{-1.78}$ & $0.20^{+0.27}_{-0.2}$ 
 & $0.88\pm0.05$  & 1.02/262 & $6.73^{+0.58}_{-0.76}$
\\
1.0-1.75 & $9.03^{+1.82}_{-1.27}$ & $0.30^{+0.25}_{-0.24}$ &
$1.08\pm0.06$  & 1.00/196 & $7.33^{+0.72}_{-0.74}$
\\
1.75-3.0 & $8.21^{+1.66}_{-1.26}$ & $0.12^{+0.25}_{-0.12}$ 
& $1.28\pm0.07$  & 0.86/189 & $7.89^{+0.81}_{-1.01}$
\\
3.0-6.0 & $5.10^{+5.68}_{-2.22}$  & 0.12(fixed) 
 & $0.73^{+0.13}_{-0.10}$  & 0.51/59 & $3.34^{+1.20}_{-1.96}$
\\
\hline
\end{tabular}
}
\end{center}
\label{tablefit}
\end{table*}

The deprojected temperature profile shows a drop in the core and a decrease 
in the outer regions (see the diamonds in the upper panel of Fig. 2), which can 
be fitted by the following formula (Xue et al. 2004):
\begin{equation}
T(r)=T_0+\frac{A}{r/r_0}exp(-\frac{(\ln r-\ln r_0)^2}{\omega}).
\end{equation}
The best fit parameters are: $A$ = $4.880\pm0.001$ keV, $r_0 = 2.494\pm0.003'$, 
$\omega$ = $2.232\pm0.004$, $T_0 = 3.084\pm0.001$ keV, $\chi^{2}_{red}=0.31$, 
and the best-fit profile is shown as the solid line in the upper panel of 
Fig. 2. From the temperature distribution, we can estimate the 
normalization-weighted temperature within $6'$, $t_{mean}=8.57^{+2.39}_{-1.56}$ 
keV, which is consistent with the results of Pointecouteau et al. (2002) and 
LaRoque et al. (2006) within the error bars and a little higher than that of 
Morandi et al. (2007). The diamonds in the bottom panel of Fig. 2 show the 
deprojected abundance distribution of RXCJ2228.6+2036.

\begin{figure}[ht]
\centerline{\psfig{file=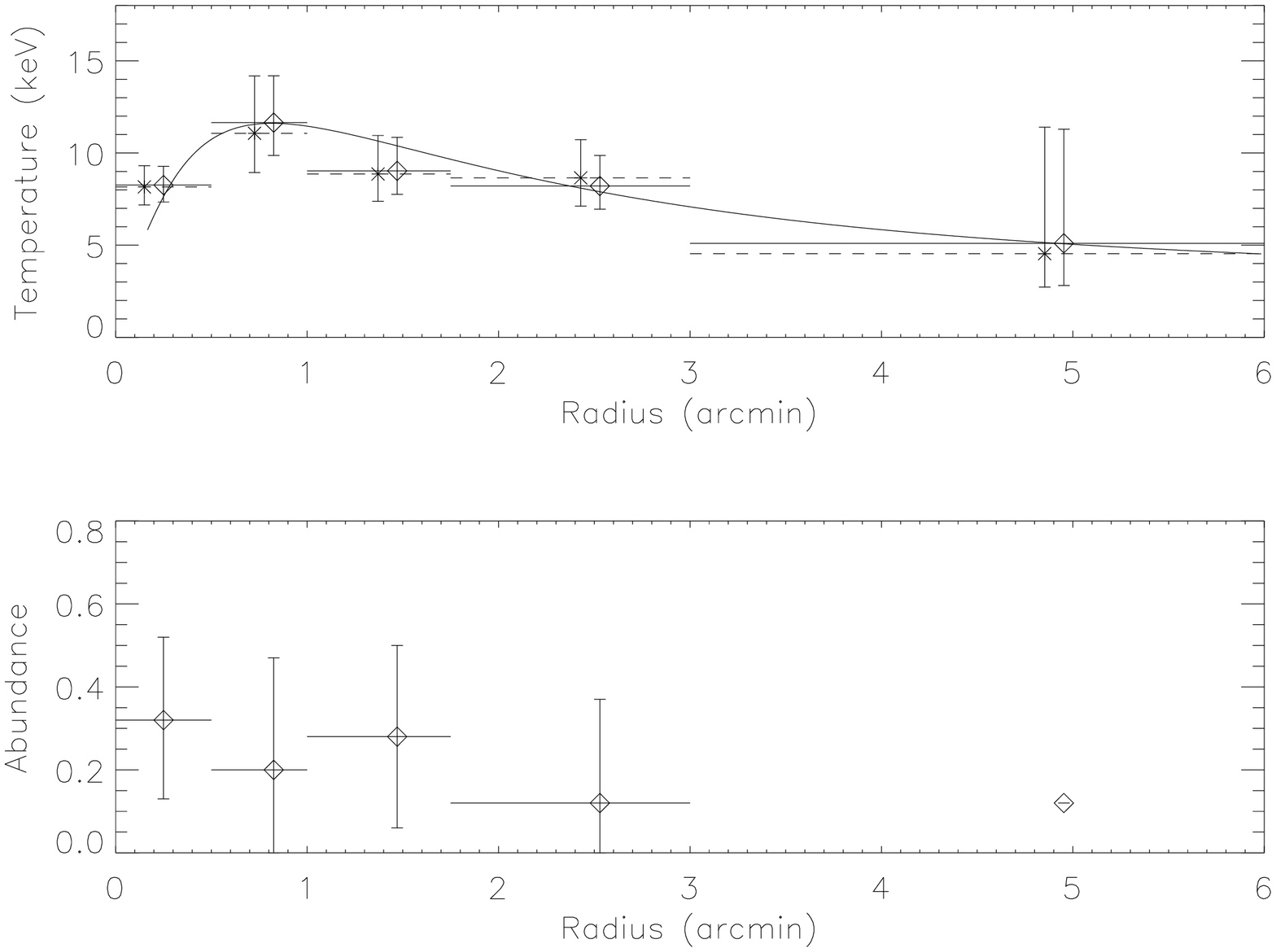,width=7cm,angle=90}}
\caption{Upper panel: Radial temperature profile of RXCJ 2228.6+2036. Diamonds: 
the deprojected temperature and the solid line is the best-fit profile. Stars: 
for the PSF-corrected temperature (see Sect. 5.1). We offset the stars $5''$ 
to the left so as to illustrate these two kinds of temperature clearly.
Bottom panel: Radial deprojected abundance of RXCJ2228.6+2036. The confidence 
level is 90\%.}
\end{figure}

%%%%%%%%%%%%%%%%%%%%%%%%%%%%%%%%%%%%%%%%%%%%%%%%%%%%%%%%%%%%%%%%%%%%%%%%%%%%%%%
\subsection{Mass deposition rate}
The temperature drop in the central part of RXCJ2228.6 +2036 might indicate the
existence of a cooling flow in the center. We thus estimate the parameters of 
the cooling flow as follows.

The cooling time $t_{cool}$ is the time scale during which the hot gas loses 
all its thermal energy, which is calculated as (e.g. Chen et al. 2007):
\begin{equation}
t_{cool}=\frac{5}{2}\frac{n_e+n_i}{n_e}\frac{k_BT}{n_H\Lambda(T)}
\end{equation}
where $\Lambda(T)$ is the cooling function of the gas, and $n_e$, $n_i$ and
$n_H$ are the number densities of the electrons, ions and hydrogen,
respectively. Here for the nearly fully ionized plasma in clusters, $n_e=1.2n_H$ 
and $n_i=1.1n_H$. The determination of $n_e$ is explained later in 
Sect. 4.1. The cooling time $t_{cool}$ of the inner two regions is given in 
Table 2. The cooling radius designates the region inside which the hot gas 
loses all its thermal energy within a cluster life time scale, usually 
using the age of the universe ($1.04\times10^{10}$ yr at $z=0.421$). The
resulting cooling radius for RXCJ2228.6+2036 is $r_{cool}=147\pm10$ kpc.

\begin{table}
\caption{The cooling time and the cooling flow rate determined with the spatial
method of the inner two regions of RXCJ2228.6+2036. The errors are at the 
68\% confidence level. $1'=332.7$ kpc.}
\begin{center}
\begin{tabular}{cccc}
\hline
 $r_1$ & $r_2$ & $t_{cool}$(yr) & $\dot{M}$(M$_{\odot}$yr$^{-1}$) 
\\
\hline
 $0'$ & $0.5'$ & $1.18\pm0.07\times10^{10}$ & $200.2\pm12.4$ 
\\
 $0.5'$ & $1'$ & $3.58\pm0.27\times10^{10}$ & $414.7\pm32.6$ 
\\
\hline
\end{tabular}
\end{center}
\label{tablefit}
\end{table}

We also fit the central spectra of pn and MOS by adding a standard cooling
flow model to the isothermal Mekal component:
\begin{eqnarray}
Model_2&=&Wabs(N_H)\times(Mekal(T,z,A,norm)+ \nonumber \\  
  && Zwabs(\Delta N_H)\times Mkcflow(\dot{M})).
\end{eqnarray}
Wabs and Mekal have been described in Sect. 3.1, Zwabs is an intrinsic
photoelectric absorption model (Morrison \& McCammon 1983), and Mkcflow is a
cooling flow model (Fabian, 1988); $\Delta N_H$ is the intrinsic absorption
and $\dot{M}$ is the rate of gas cooling out of the flow. The fitting results
show that the mass deposition rate in the central region $r<0.5'$ is
$14.0^{+64.0}_{-14.0}$ M$_{\odot}$yr$^{-1}$ (see Table 3), so within $r_{cool}$
$\dot{M}=12_{-12}^{+56}$ M$_{\odot}$yr$^{-1}$. Fig. 3(b) presents the central 
spectra fitted by this model.

\begin{table*}
\caption{The best-fit parameters for the central region of RXCJ2228.6+2036 by 
$Model_2$. The $lowT$ is fixed on 0.01 keV and $1'=332.7$ kpc. The errors 
are at the 90\% confidence level.}
\begin{center}
\begin{tabular}{ccccccccc}
\hline
r & $T_{mekal}$ & low$T_{cf}$ & high$T_{cf}$ & A 
  & norm & $\dot{M}$ & $\Delta n_H$ & $\chi^{2}_{\mathrm{red}}$/$dof$ 
\\
  &(keV) & (keV) & (keV)& (solar) & ($10^{-3}$cm$^{-5}$) & 
(M$_{\odot}$) & ($10^{22}$cm$^{-2}$) &   
\\
\hline
 $0'-0.5'$ & $8.54^{+2.27}_{-1.18}$ & 0.01 (fix) & =$T_{mekal}$ & 
 $0.32\pm0.19$ & $0.70^{+0.05}_{-0.12}$ & $14.0^{+64.0}_{-14.0}$  & 
 0.0(fix) &  1.12/250 
\\
\hline
\end{tabular}
\end{center}
\label{tablefit}
\end{table*}

\begin{figure*}
\hbox{\psfig{figure=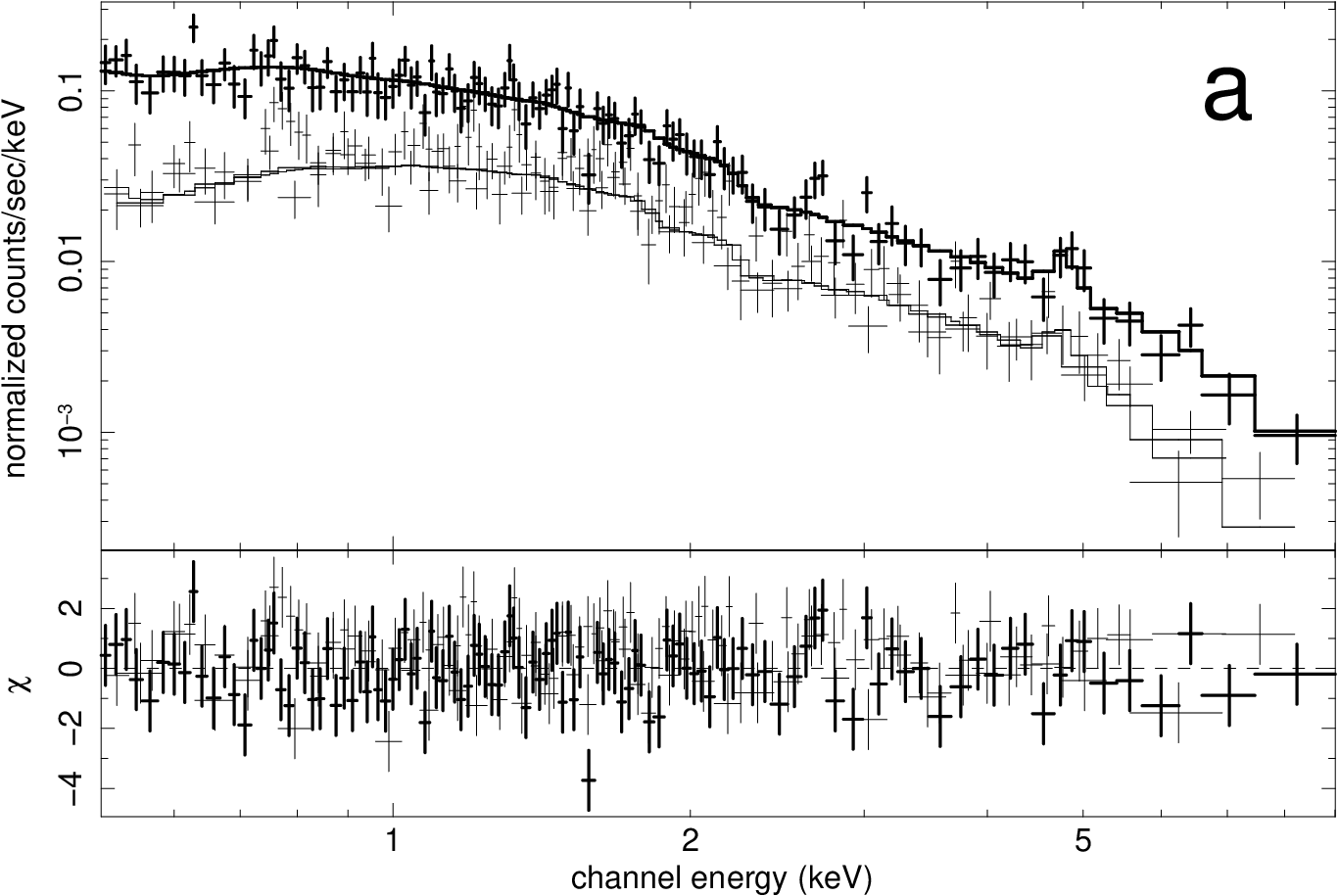,height=8.5cm,angle=-90}
\psfig{figure=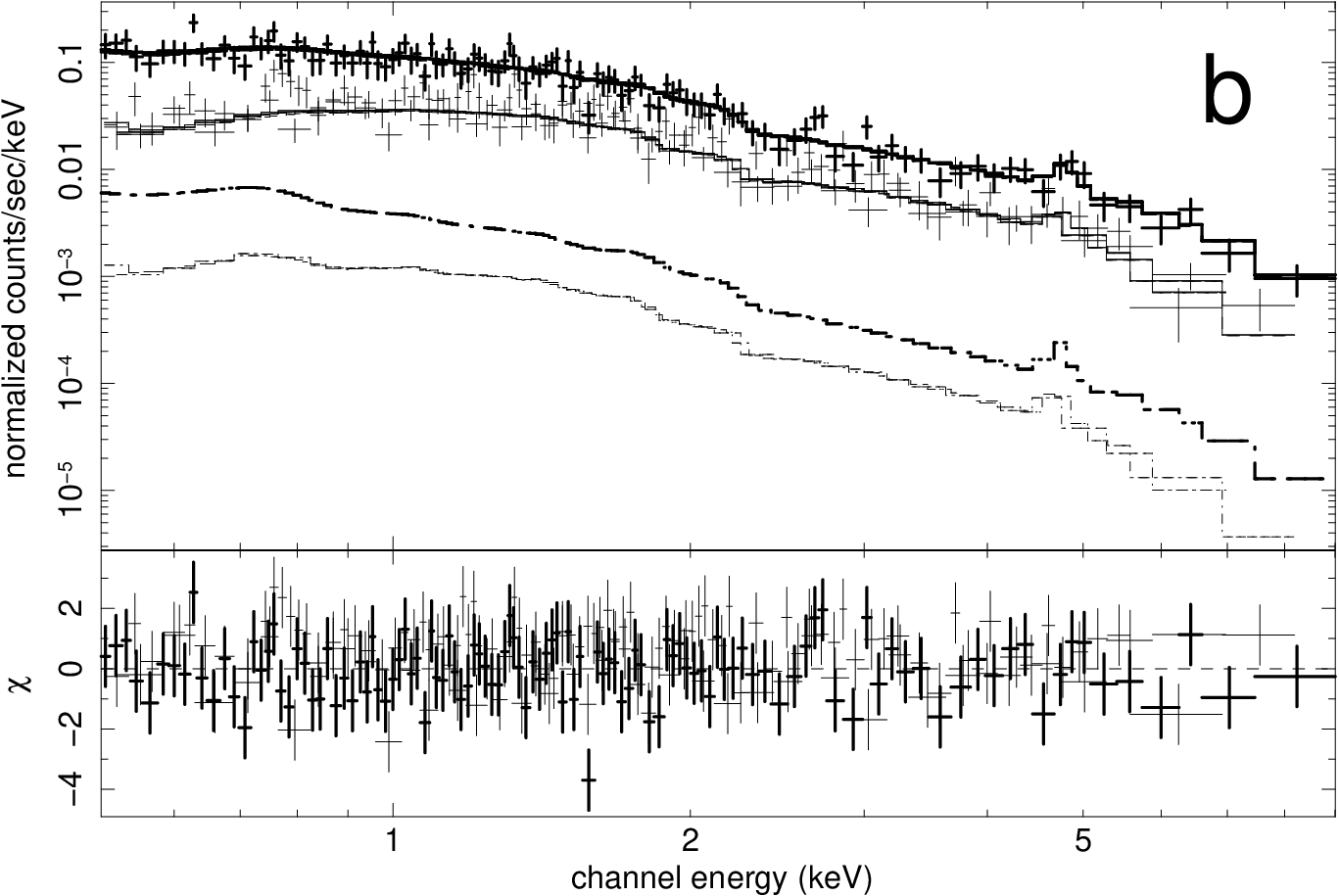,height=8.5cm,angle=-90}}
\hbox{}
\caption {The spectra of the central region (r $<0.5'$) for joint fit of pn
(bold crosses) and MOS (faint crosses) of RXCJ2228.6+2036. {\bf a)} fitted 
by a single-temperature model; {\bf b)} fitted by a cooling flow model with 
an isothermal Mekal component. In b) we plot the isothermal and the cooling 
flow components respectively, and the lower lines below the crosses represent 
the multiphase components of pn (bold line) and MOS (faint lines), which 
show that the multiphase components only contribute a little to the emissions.}
\end{figure*}

%%%%%%%%%%%%%%%%%%%%%%%%%%%%%%%%%%%%%%%%%%%%%%%%%%%%%%%%%%%%%%%%%%%%%%%%%%%%%%%
\section{Mass determination}
\subsection{Electron density}
We divided the $r<6^{\prime}$ region into 13 annuli centered on the emission
peak, where the width of each annular region is determined to obtain at least
2000 total counts in each annulus region in 0.5-10 keV. After the vignetting 
correction and the double-background subtraction, the surface brightness 
profile for RXCJ2228.6+2036, $S(r)$, is derived, which can be fitted by a 
double-$\beta$ model (as Eq. 9) convolved with the PSF matrices (Ghizzardi 
2001) to correct for the PSF effect. Using the deprojection technique described 
in Sect. 2.3, we deproject the double-$\beta$ model (PSF corrected) and obtain 
the count rate of each corresponding shell, $Ctr(i)$. Since the temperature and 
abundance profiles are known, giving $T(i)$, $A(i)$ and $Ctr(i)$ in the $i$th 
shell, we can derive the corresponding $norm(i)$ in XSPEC with $Model_1$. The 
radial electron density $n_e$ of each region can be determined from Eq. (8), 
shown as stars in Fig. 4.

\begin{eqnarray}
norm(i)= 10^{-14}/(4\pi D^2)\cdot \int n_e n_H dV.
\end{eqnarray}

We fit the derived electron density (the stars in Fig. 4) with the 
double-$\beta$ model (Chen et al. 2003):
\begin{equation}
n_e(r)=n_{01} {\left(1+{(\frac{r}{r_{c1}})}^2\right)}^{-{\frac{3}{2}}\beta_1}+n_{02}
{\left(1+{(\frac{r}{r_{c2}})}^2\right)}^{-{\frac{3}{2}}\beta_2},
\end{equation}
the best-fit parameters are: $n_{01}$ = $0.0027\pm0.0002$ cm$^{-3}$, 
$r_{c1}$ = $2.5480\pm0.0002$ arcmin, $\beta_1$ = $1.4031\pm0.0001$, 
$n_{02}$ = $0.0109\pm0.0001$ cm$^{-3}$, $r_{c2}$ = $0.6547\pm0.0004$ arcmin, 
$\beta_2$ = $1.5474\pm0.0001$, $\chi^{2}_{red}=0.71$, $dof$=7. 
And the best-fit profile is shown as the solid line in Fig. 4.

\begin{figure}[ht]
\centerline{\psfig{file=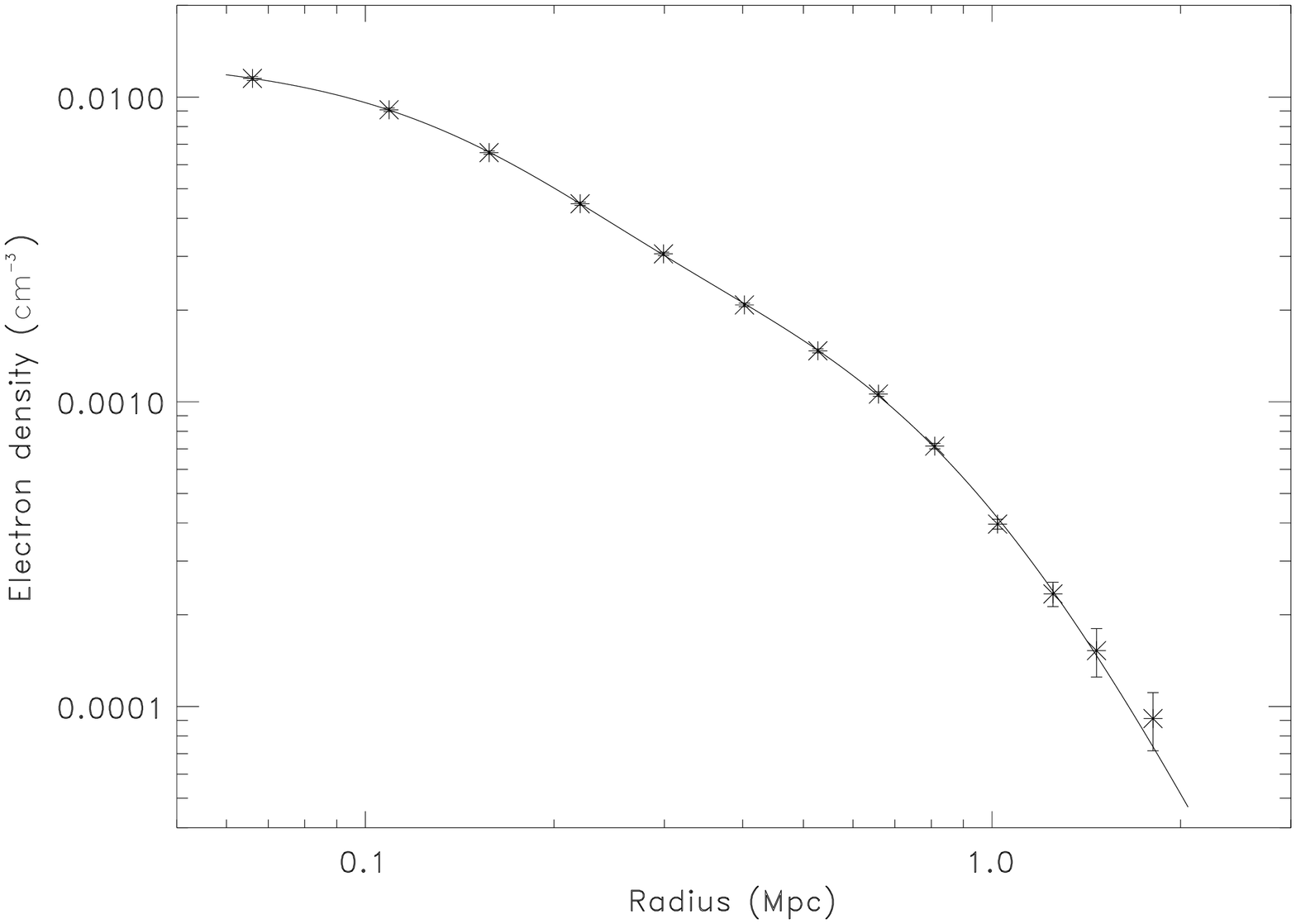,width=6.5cm,angle=90}}
\caption{The deprojected electron density profile of RXCJ2228.6+2036 after PSF
correction. The error bars are at the 68\% confidence level. The solid line 
is the best-fit profile from a double-$\beta$ model fitting.}
\end{figure}

\subsection{Total mass}
Once we have derived the radial temperature profile $T(r)$ and electron
density profile $n_e(r)$ for RXCJ2228.6+2036, the integrated total mass of this
cluster at radius $r$ can be calculated under the assumptions of hydrostatic
equilibrium and spherical symmetry by the following equation (Fabricant et al.
1980):
\begin{equation}
M_{tot}(<r)=-\frac{k_B T r^2}{G\mu
m_p}\left[\frac{d(\ln{n_e})}{dr}+\frac{d(\ln{T})}{dr}\right],
\end{equation}
here $k_B$ is the Boltzmann constant, $G$ is the gravitational constant and
$\mu$ is the mean molecular weight in units of the proton mass $m_p$ (we assume 
$\mu=0.6$ in this work). The mass uncertainties are obtained from the
uncertainties of the temperature and the electron density calculated by 
Monte-Carlo simulations. We obtained 250 redistributions of the parameterized
temperature and electron density profiles by fitting to the data points varied 
within the Gaussian error bars of the measurements. The uncertainties of all 
the other properties of RXCJ2228.6+2036 are also calculated from the 250 
simulated clusters.

Then, we derived the total mass profile of RXCJ2228.6 +2036, shown in Fig. 5, and 
$M_{tot}=(1.36\pm0.51)\times10^{15}$ M$_{\odot}$ at $6'$ at the 68\% confidence 
level. 

\begin{figure}[ht]
\centerline{\psfig{file=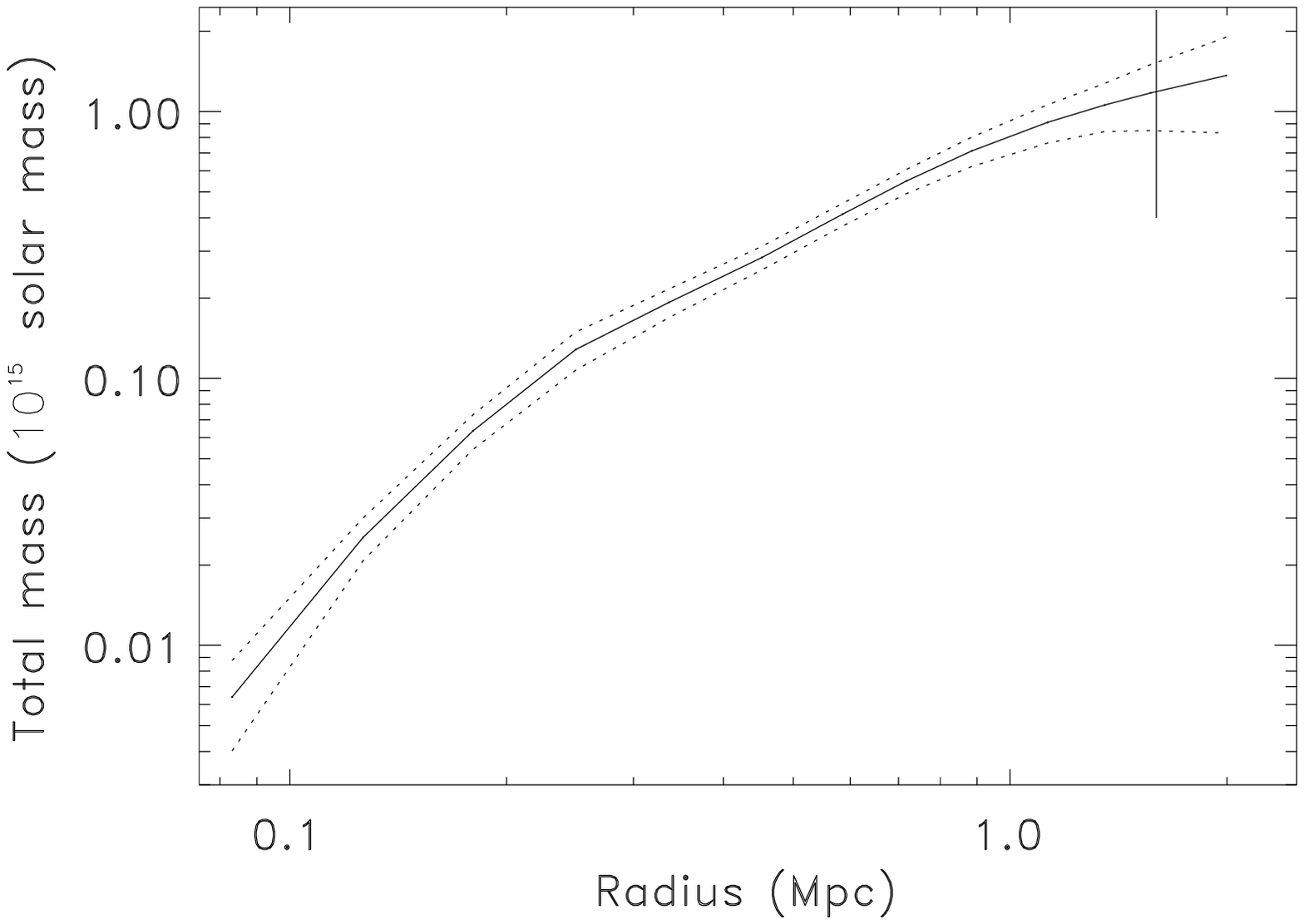,width=9cm}}
\caption{The total mass profile of RXCJ2228.6+2036, and the error bars (dotted 
lines) are at the 68\% confidence level. The vertical line indicates
$r_{500}=1.61\pm0.16$ Mpc.}
\end{figure}

A physically meaningful radius for the mass measurement is defined as
$r_{500}$, the radius within which the mean gravitational mass density
$<\rho_{tot}>=500\rho_c$, where $\rho_c=3H^2/(8\pi G)$ is the critical cosmic
matter density. For our calculations, we use the value of $\rho_c$ at the 
cluster redshift, i.e., $\rho_c=9.2\times10^{-30}$ g cm$^{-3}$. This radius is 
still well covered by the observations. From the mass profile we derive 
$r_{500}=1.61\pm0.16$ Mpc for RXCJ2228.6+2036, corresponds to $4.8'\pm0.5'$, 
and the total mass within it is about $M_{500}=(1.19\pm0.35)\times10^{15}$ 
M$_{\odot}$. The mass derived from {\it Chandra} data by Morandi et al. (2007) 
is $M_{tot}=(4.90\pm4.35)\times10^{14}$ M$_{\odot}$ within 
$r_{500}=1033 \pm 464$ kpc, still consistent with ours within the error bars. 
In our analysis, the extrapolated value of $r_{vir}=2.48\pm0.38$ Mpc 
$=7.5'\pm1.1'$ and $M_{vir}=(1.55\pm0.72)\times10^{15}$ M$_{\odot}$, which 
agrees with what Pointecouteau et al. (2002) derived.

\subsection{gas mass and gas mass fraction}
In galaxy clusters, gas is an important component involving complex physics.
From the electron density we calculate the gas mass and the gas mass fraction
defined as $f_{gas}(r)=M_{gas}(r)/M_{tot}(r)$.  Fig. 6 shows the gas mass
fraction profile of RXCJ2228.6+2036. The gas mass fraction at $6'$ is
$f_{gas}=0.17\pm0.06$, consistent within the error bars with the results of
LaRoque et al. (2006) from the {\it Chandra} X-ray and OVRO/BIMA
interferometric SZ effect measurements: $f_{gas} = 0.138 \pm 0.009$ from the
X-ray data and $f_{gas} = 0.188^{+0.035}_{-0.031}$ from the SZ data. It also
agrees with the WMAP measured baryon fraction of the Universe $f_b =
\Omega_b/\Omega_m = 0.166$ (Spergel et al. 2003).

\begin{figure}[ht]
\centerline{\psfig{file=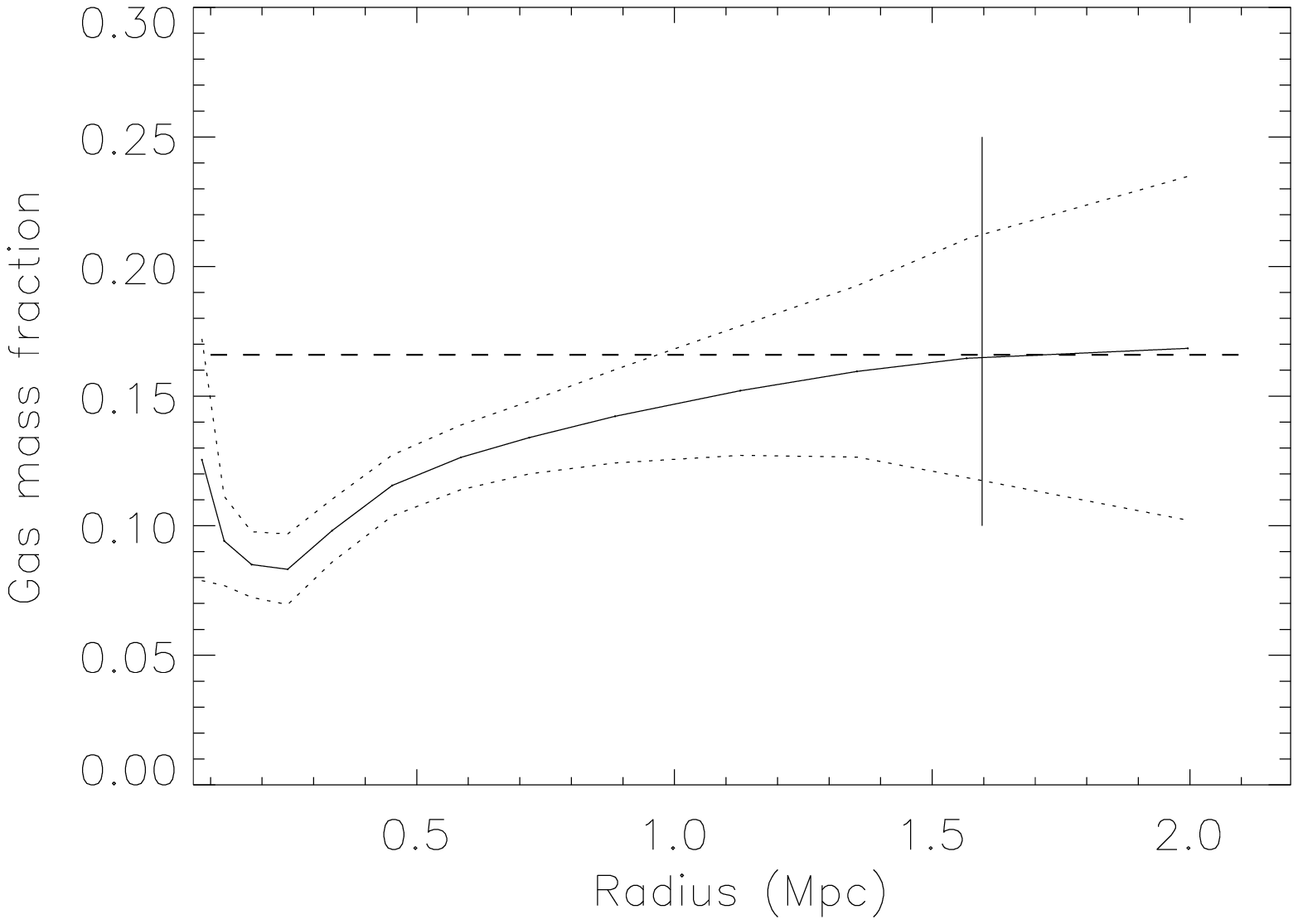,width=9cm}}
\caption{The gas mass fraction profile of RXCJ2228.6+2036. The dashed horizontal 
line indicates the WMAP measurement of $f_b=\Omega_b/\Omega_m=0.166$ (Spergel 
et al. 2003) and the vertical line indicates $r_{500}=1.61\pm0.16$ Mpc. The 
error bars (dotted lines) are at the 68\% confidence level.}
\end{figure}

%%%%%%%%%%%%%%%%%%%%%%%%%%%%%%%%%%%%%%%%%%%%%%%%%%%%%%%%%%%%%%%%%%%%%%%%%%%%%%%
\section{Discussion}
\subsection{PSF-corrected spectra}
The spatially resolved spectral analysis is affected by the PSF.
To correct for the PSF effect, we first calculate the redistribution matrix,
$F_{ij}$, which is the fractional contribution in the $i$th ring coming from
the $j$th ring (Pratt \& Arnaud 2002). We get this redistribution from our 
best fitted double-$\beta$ model of electron density (converted to emission 
measure profile) and the PSF matrices (Ghizzardi 2001). Since we have divided 
our cluster into 5 regions, we can obtain the fractional contribution in each 
ring coming from all the bins, and in Fig. 7, we plot the contribution from 
the bin, all inner and outer bins.

\begin{figure}[ht]
\centerline{\psfig{file=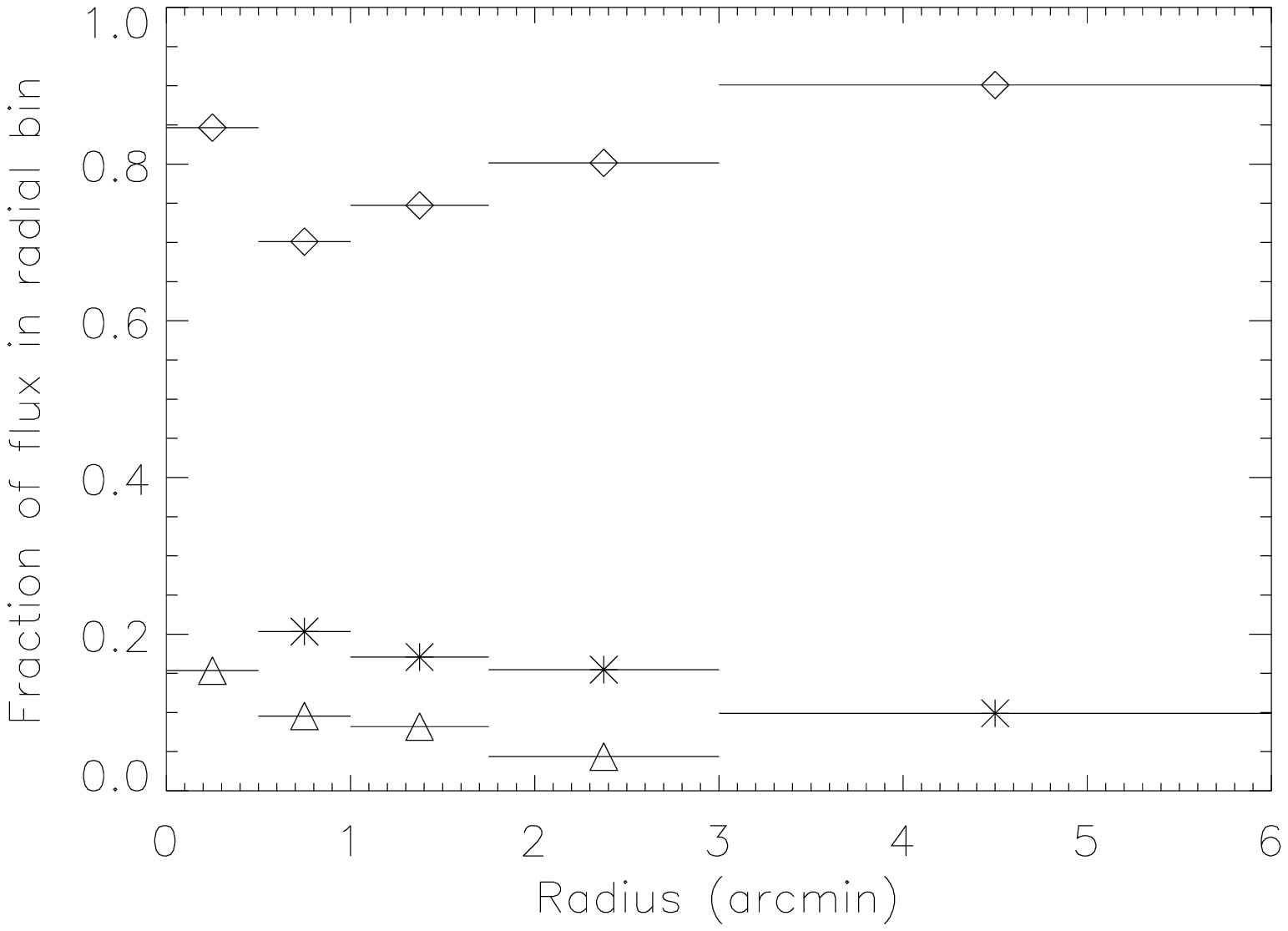,width=9cm}}
\caption{Redistributions due to the {\it XMM-Newton} PSF: the diamonds
represent the contribution coming from the bin, the stars from the inner bins
and the triangles from the outer bins.}
\end{figure}

Here, $O_i$ is the observed spectrum of the $i$th ring after a double-background 
subtraction, $S_i$ is the spectrum without PSF effect, so for our cluster 
RXCJ2228.6+2036, we have:

\begin{eqnarray}
\left(\begin{array}{ccccc}
F_{11}&F_{12}&F_{13}&F_{14}&F_{15}\\
F_{21}&F_{22}&F_{23}&F_{24}&F_{25}\\
F_{31}&F_{32}&F_{33}&F_{34}&F_{35}\\
F_{41}&F_{42}&F_{43}&F_{44}&F_{45}\\
F_{51}&F_{52}&F_{53}&F_{54}&F_{55}
\end{array}\right) 
\cdot 
\left(\begin{array}{c}
S_1\\
S_2\\
S_3\\
S_4\\
S_5
\end{array}\right) 
=
\left(\begin{array}{c}
O_1\\
O_2\\
O_3\\
O_4\\
O_5
\end{array}\right).
\end{eqnarray}
From this function we can derive $S_i$, which indicates the spectra after
the PSF correction. Then, we deproject these spectra $S_i$ by the deprojection
technique described in Sect. 2.3, and thus derive the PSF corrected deprojected
spectra. Fitting these spectra with $Model_1$, we obtained the PSF-corrected
deprojected temperatures, shown as stars in the upper panel of Fig. 2.

We find that the PSF-corrected temperatures agree with the measurements
without PSF-correction. This may be due to the broad width of the regions we
chose. However it should be noted here that these spectral fits are not as good
as those in Sect. 3.1 because the PSF-correction procedure introduces
significant uncertainties, mainly due to the inversion process (see Eq. 11).

\subsection{Gas pressure and comparison with the SZ data}
With the temperature profile $T(r)$ and electron density profile $n_e(r)$, we
can derive the gas pressure profile of RXCJ2228.6+2036 as:
\begin{equation}
P(r)=n_e(r)k_BT(r). 
\end{equation}
In terms of observables, the gas pressure can be checked against the SZ effect 
(Sunyaev \& Zel'dovich 1972) coming from the cluster. Indeed the SZ effect is
directly proportional to the integrated pressure over the line of sight:
\begin{eqnarray}
S_{SZ}(r)&=&\frac{\sigma_T}{m_ec^2}
\int{k_BT(r)n_e(r)dr} \cdot f_{SZ}(v,T) \nonumber \\
&=&\frac{\sigma_T}{m_ec^2} \int{P(r)dr} \cdot f_{SZ}(v,T),
\end{eqnarray}
where $k_B$, $m_e$, $c$ and $\sigma_T$ are the Boltzmann constant, the electron 
mass, the speed of light and the Thomson cross-section; $f_{SZ(v,T)}$ 
represents the SZ spectral shape (including the relativistic corrections
as computed by Pointecouteau et al. 1998). 

Here we integrated the gas pressure of RXCJ2228.6 +2036 using Eq. (13), convolved
with the PSF of the 45 m radio telescope NRO (the beam size at 21 GHz: 
$\theta_{FWHM}\sim80$ arcsec) and then compared it with the SZ radial
profile (Pointecouteau et al. 2002) in Fig. 8. The diamonds are from the SZ
data and the solid line represents our result. We found a remarkable agreement
within the error bars.

\begin{figure}[ht]
\centerline{\psfig{file=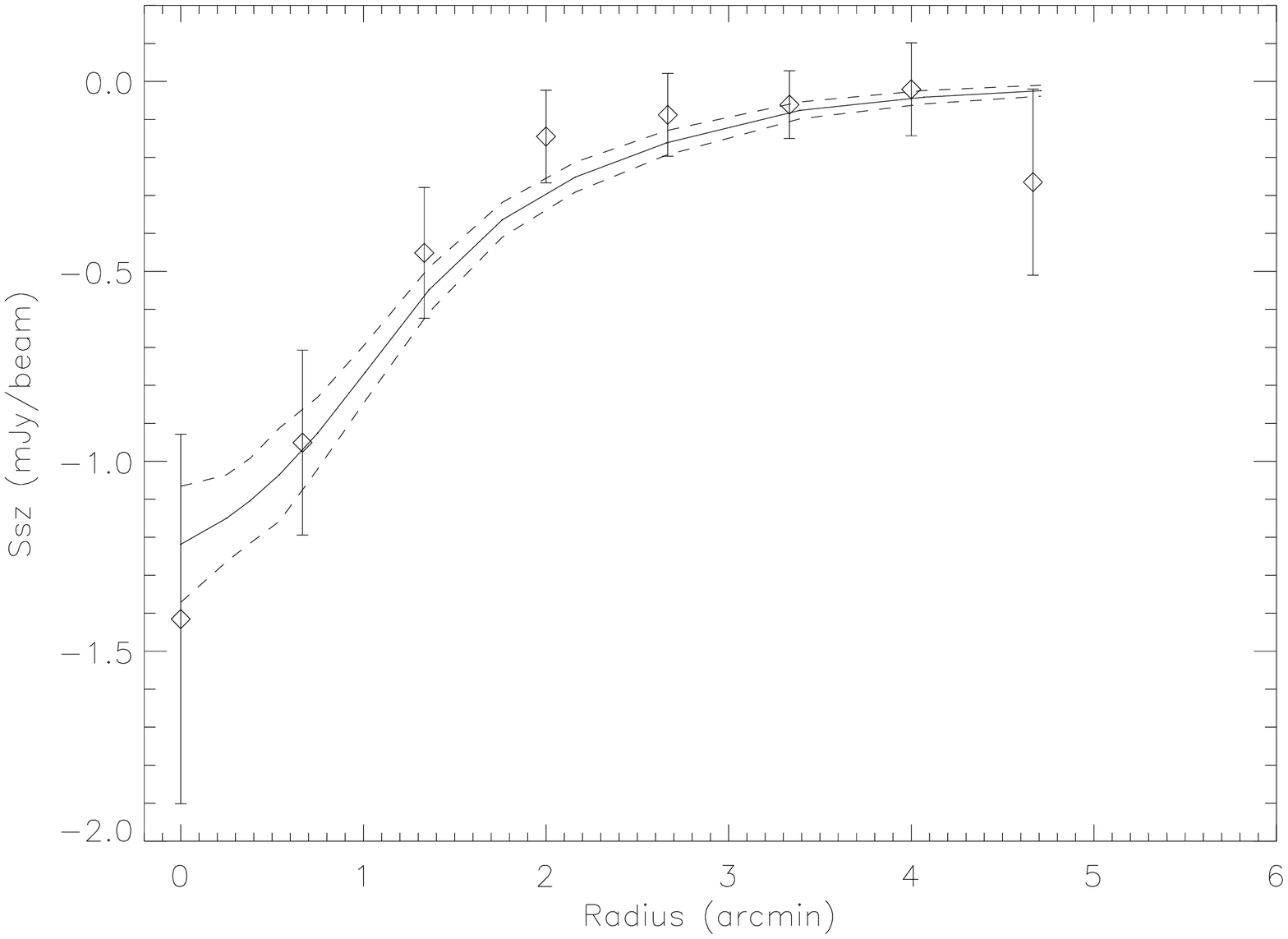,width=6.5cm,angle=90}}
\caption{The integrated X-ray pressure profile of RXCJ2228.6 +2036 convolved
with the PSF of the SZ telescope (the solid line) and compared with the SZ 
radial profile (the diamonds) derived by Pointecouteau et al. (2002). All 
the errors are at the 68\% confidence level.}
\end{figure}

The good agreement of the X-ray and SZ surface brightness profile in Fig. 8 may
lend itself to test biases in the derivation of the pressure profile from the
X-ray data. The most interesting aspects concern a bias in the temperature
measurement in the presence of a multitemperature ICM (e.g. Mazzotta et al.
2004, Vikhlinin 2006) and the overestimate of the gas density due to the
enhancement of the surface brightness if the gas is clumpy. In the following we
will investigate how these two effects are modifying the comparison of the X-ray
and SZ data.

If we assume that the ICM is in rough pressure equilibrium the two bias effects
on the temperature and the density are actually linked (for $n_e \cdot T=$
constant). While local unresolved density inhomogeneities will lead to an
overestimation of the density and the prediction of a too high SZ-signal, the
temperature of a clumpy medium will be underestimated compared to a mass average
and result in an underprediction of the SZ signal. So both effects are at least
partly compensating each other in our study.

Quantitatively the overestimation of the gas density is given by:
\begin{equation}
C'=\frac{<n_e^2>}{<n_e>^2}, 
\end{equation}
where the overestimation factor is $C=\sqrt{C'}$.
To quantify the underestimation of the temperature for this hot cluster, {\bf we 
can approximately use the approach of Mazzotta et al. (2004) (Eq.14) which 
yields:
\begin{equation}
\frac{T_{sl}}{T}=R=\frac{<n_e><n_e^{1.75}>}{<n_e^{2.75}>}, 
\end{equation}
where $T_{sl}$ is a good approximation of the spectroscopic temperature as would
be obtained from data analysis of {\it Chandra} and {\it XMM-Newton} observations
for a multi-temperature plasma, and $T$ the mass weighted mean.}

As an example we calculate these effects for a homogeneous distribution with a
lower and higher cutoff of, $T_1=<T>-\varepsilon$ and $T_2=<T>+\varepsilon$,
respectively. A more general distribution can also be seen as a superposition 
of many of these `top-hat' distributions. Fig. 9 shows the enhancement factors
$C$ and $R$ as a function of the distribution width parameter, $\varepsilon$. 
We note that the two effects don't cancel each other, but the effect of the
temperature underestimate is about 2-3 times larger than the overestimat-e due to
clumpiness. Still the overall effect is not dramatic and does therefore not
provide a very good diagnostics. Even for a broad temperature distribution with
$\varepsilon/<T> \approx 0.75$ for example, covering a temperature range (from
the lower temperature to the higher temperature) of a factor of 7, we obtain an 
SZE underestimate of about {\bf 30\% and a gas mass overestimate of about 18\%.}

\begin{figure}[ht]
\centerline{\psfig{file=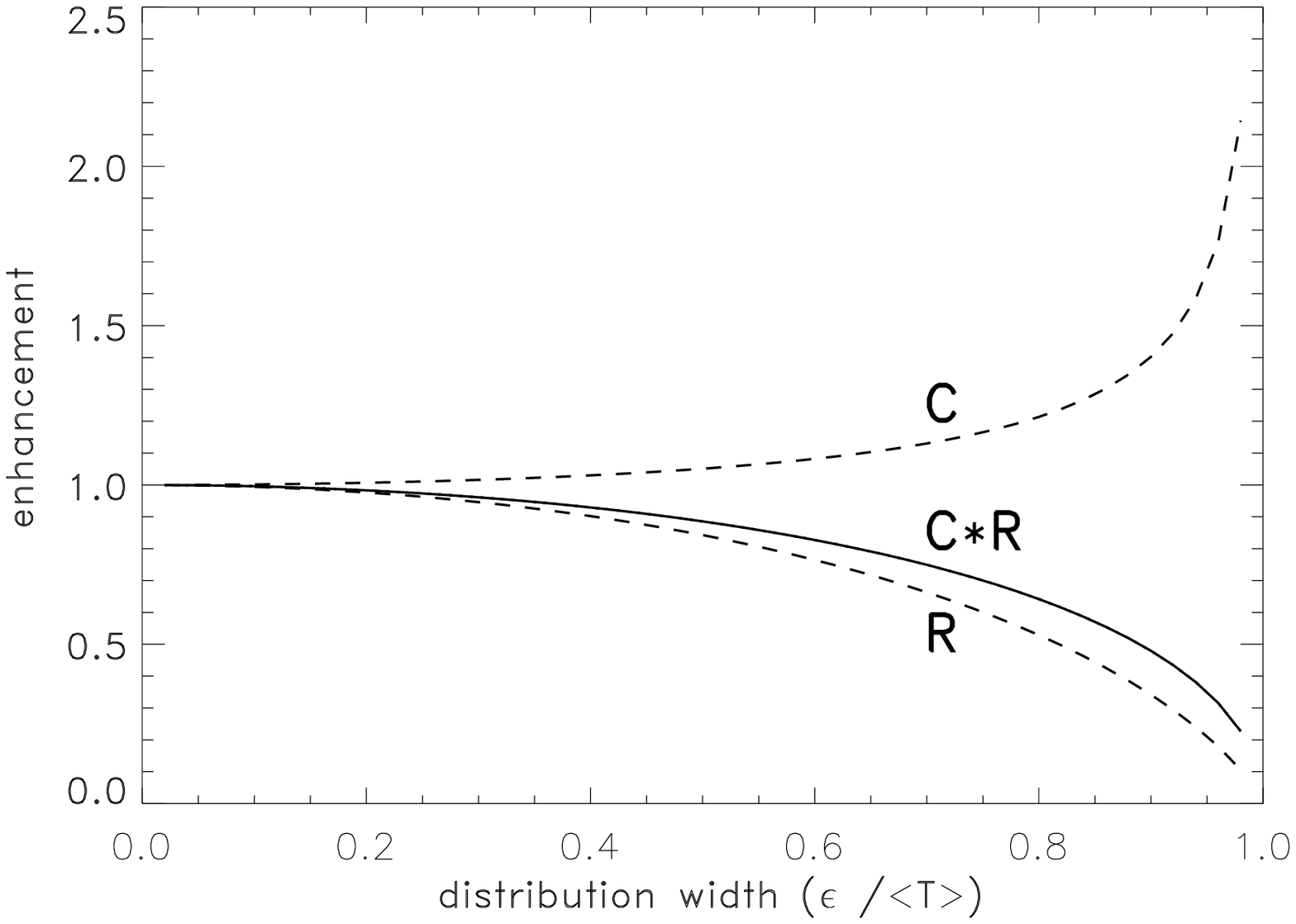,width=8.5cm}}
\caption{Overestimation factor of the gas density $C$ and underestimation of 
the spectroscopic like temperature versus the mass weighted temperature $R$ as 
a function of the width of a homogeneous temperature distribution in the 
presence of pressure equilibrium. The combined effect $C\times R$ is for the 
underestimation of SZE. For the definition of the parameters see the text.}
\end{figure}

This has also implications on the mass measurement. While the pressure profile
and its derivative can be directly taken from the SZ-profile, we still require
an independent absolute temperature measurement for the normalization of the
mass profile. The above calculation shows now, that we don't obtain very precise
new information on a possibly low biased temperature due to a multiphase ICM 
from having simultaneous X-ray and SZ observations. In the above example a
temperature and mass underestimation of {\bf 40\% is only indicated by an SZ
deviation of 30\%.}

\subsection{Gas entropy and the $S-T$ relation}
Following Ponman et al. (1999), we defined the entropy of the gas in clusters
as:
\begin{equation}
S(r)=\frac{T(r)}{n_e(r)^{\frac{2}{3}}}. 
\end{equation}
This entropy corresponds to the heat supplied per particle for a given reference
density. Fig. 10 shows the entropy distribution as a function of radius, where 
the diamonds represent the entropy obtained from the spectra fitting results 
and the solid line from the best fitted $T(r)$ and $n_e(r)$ profiles.

\begin{figure}[ht]
\centerline{\psfig{file=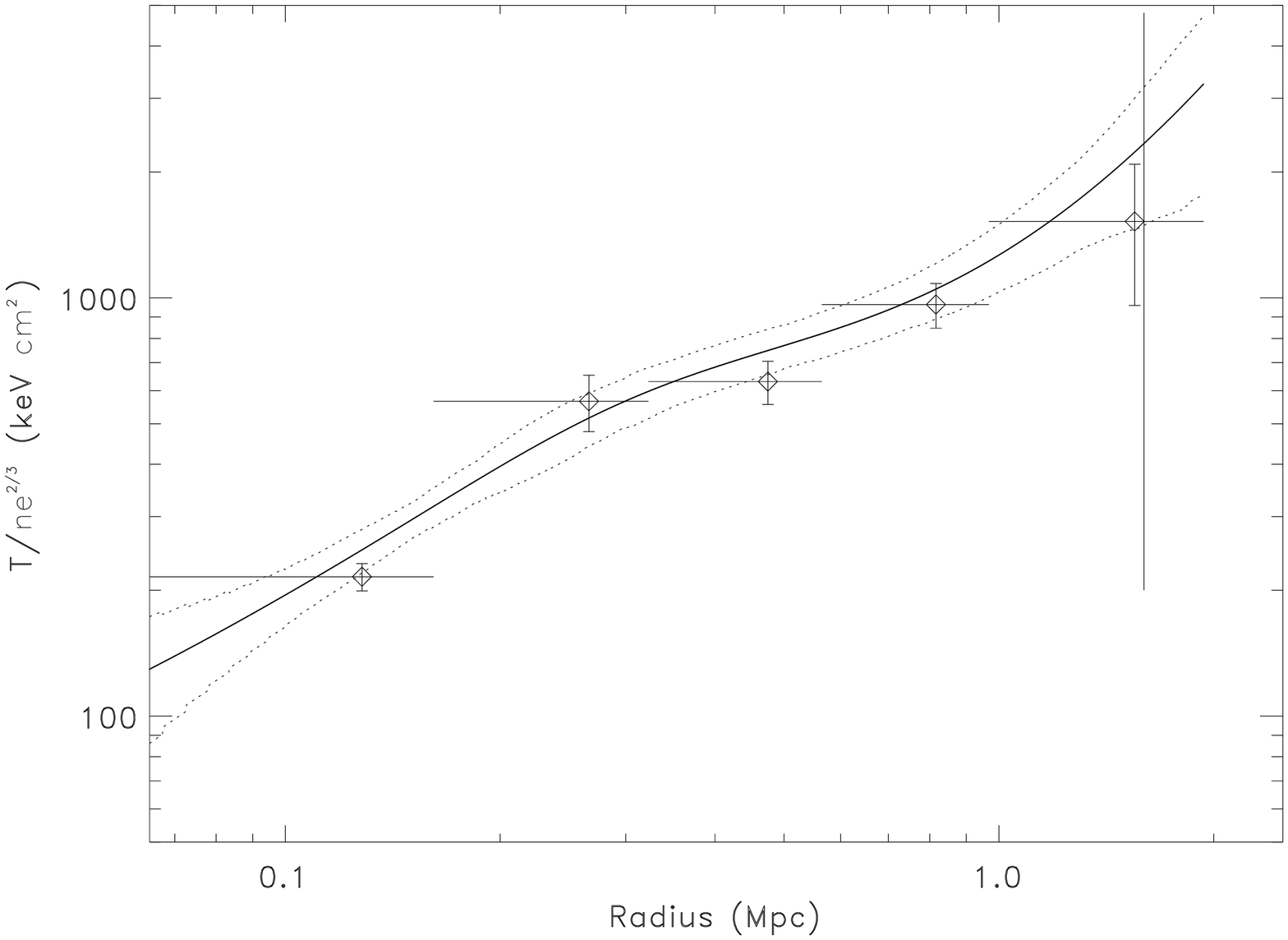,width=6.5cm,angle=90}}
\caption{The entropy distribution of RXCJ2228.6+2036. The diamonds represent
the entropy derived from the spectral fitting results and the solid line from 
the best fitted $T(r)$ and $n_e(r)$ profiles. The vertical line shows 
$r_{500}=1.61\pm0.16$ Mpc. The error bars are at the 68\% confidence level.}
\end{figure}

Pratt et al. (2006) have shown for the $S-T$ relation measured from a sample 
of 10 local and relaxed clusters observed by {\it XMM-Newton}, that 
$S_{0.3} \propto T_X^{0.64}$, where
$S_{0.3}$ means the entropy at $0.3r_{200}$ and $T_X$ is the mean temperature 
in the region of $0.1r_{200}<r<0.5r_{200}$. For RXCJ2228.6+2036, 
$S_{0.3r200}=959\pm130$ keV cm$^2$ and $T_X=8.91^{+1.91}_{-1.33}$ keV. 
The $S_{0.3r200}$ versus $T_{X}$ for RXCJ2228.6+2036 is plotted on the $S-T$ 
relation derived by Pratt et al. (2006), shown in Fig. 11. The diamonds and
the best fit line (the solid line) are from Pratt et al. (2006), and the star 
indicates the result of RXCJ2228.6+2036. It shows that our entropy value for
RXCJ2228.6+2036 is consistent (within the 1$\sigma$ error bars) with Pratt et
al. (2006) $S-T$ relation at 0.3$r_{200}$, once corrected for the expected
evolution in a self-similar scenario of structure formation.

\begin{figure}[ht]
\centerline{\psfig{file=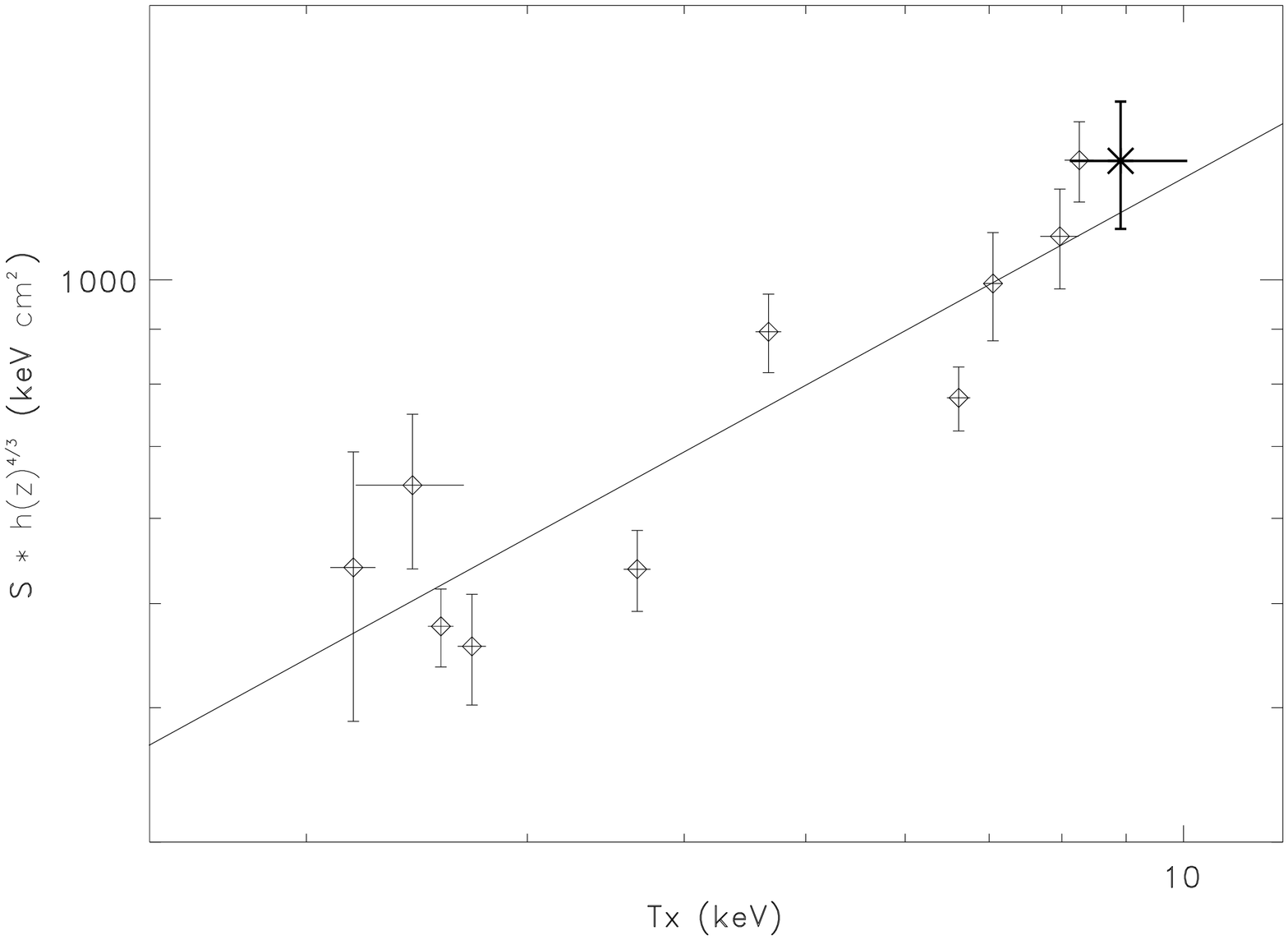,width=6.5cm,angle=90}}
\caption{Comparison of the present result with the $S-T$ relation of Pratt et 
al. (2006). The star indicates the result of RXCJ2228.6+2036, and the diamonds 
and the best-fitted $S-T$ relation line (the solid line) come from Pratt et al. 
(2006). Here $h(z)=[0.3(1+z)^3+0.7]^{\frac{1}{2}}$. The error bars of the star 
are at the 68\% confidence level.}
\end{figure}

\subsection{$M-T$ and $L-T$ relations}
From the above analysis, we derived the temperature, mass and X-ray luminosity
(see Table 1) of RXCJ2228.6+2036: within $r_{500}=4.8'$,
$M_{500}=(1.19\pm0.35)\times10^{15}$ M$_{\odot}$,
$T_{500}=8.92^{+1.78}_{-1.32}$ keV and $L_{bol,500}=28.83^{+3.69}_{-4.78} 
\times 10^{44}$ erg s$^{-1}$. Therefore we can compare RXCJ2228.6+2036 to the 
empirical scaling relations for massive galaxy clusters, e.g. $M_{500}-T_{500}$ 
and $L_{bol,500}-T_{500}$ derived from {\it XMM-Newton} data, such as in Kotov 
\& Vikhlinin (2005) based on 10 clusters at $z=0.4-0.7$, Arnaud et al. (2005) 
based on 10 nearby clusters ($z<0.15$) and Zhang et al. (2008) based on 37
LoCuSS clusters at $z\sim 0.2$. The comparison is shown in Figs.12-13. The 
diamonds and the solid line are from Kotov \& Vikhlinin (2005), the triangles 
and the dashed line from Arnaud et al. (2005), the dotted line from Zhang 
et al. (2008), and the star indicates the result of RXCJ2228.6+2036. 

It shows that our result is consistent with any of these previous studies 
within the scatter of the relations. The agreement of our $L-T$ relation with 
that of Kotov \& Vikhlinin (2005, with objects in the same redshift range as 
ours) is remarkable, in particular as X-ray luminosity with its square
dependence on density is a parameter which is very sensitive to morphological
disturbances and which thus generally shows a large scatter.

\begin{figure}[ht]
\centerline{\psfig{file=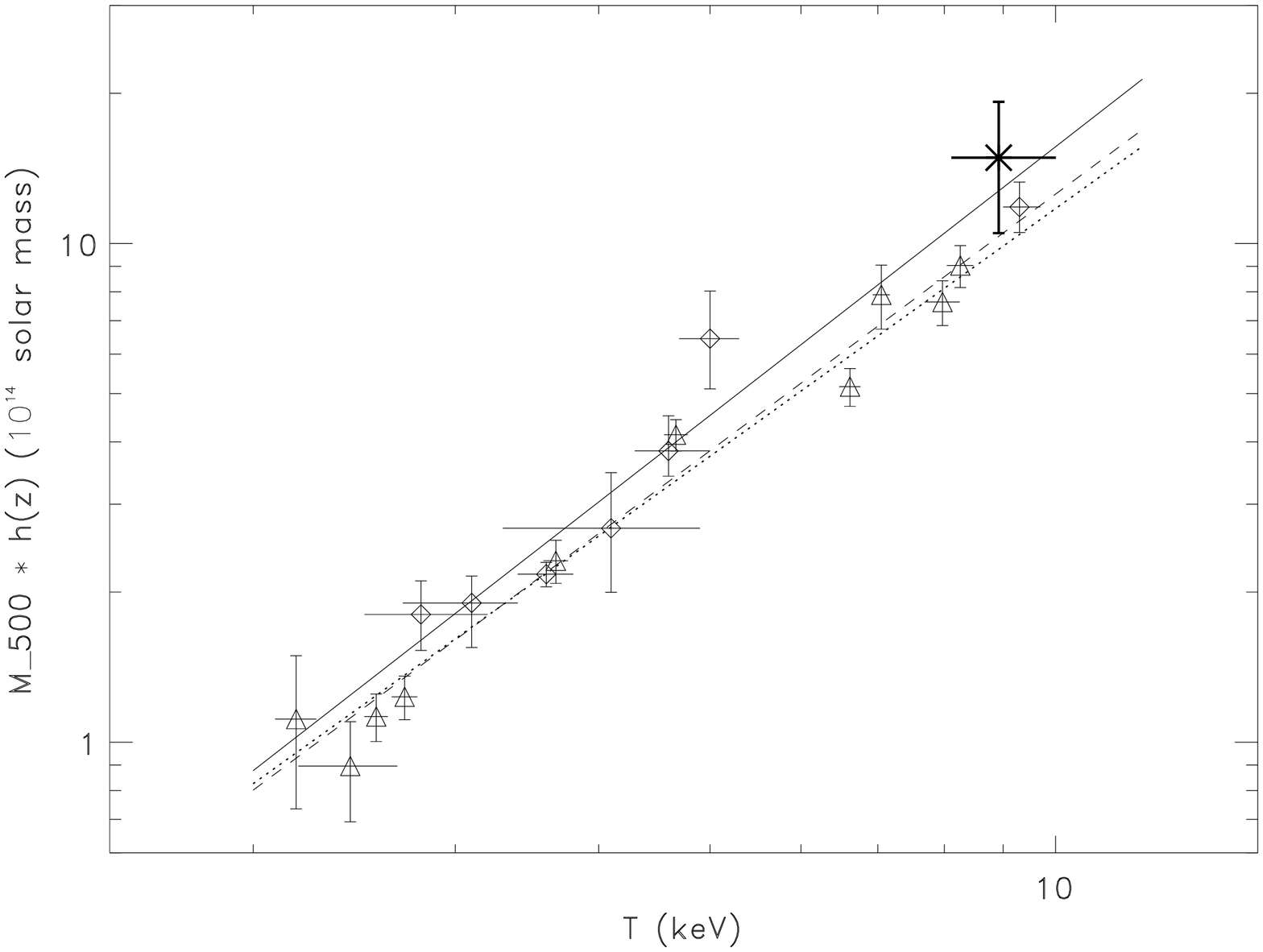,width=6.5cm,angle=90}}
\caption{Comparison of the present result with literature $M-T$ relations. The 
star indicates the result of RXCJ2228.6+2036, the diamonds and their best-fitted 
$M-T$ relation line (the solid line) come from Kotov \& Vikhlinin (2005), the 
triangles and the dashed line from Arnaud et al. (2005), and the dotted line 
from Zhang et al. (2008). Here $h(z)=[0.3(1+z)^3+0.7]^{\frac{1}{2}}$. The 
error bars of the star are at the 68\% confidence level.}
\end{figure}
\begin{figure}[ht]
\centerline{\psfig{file=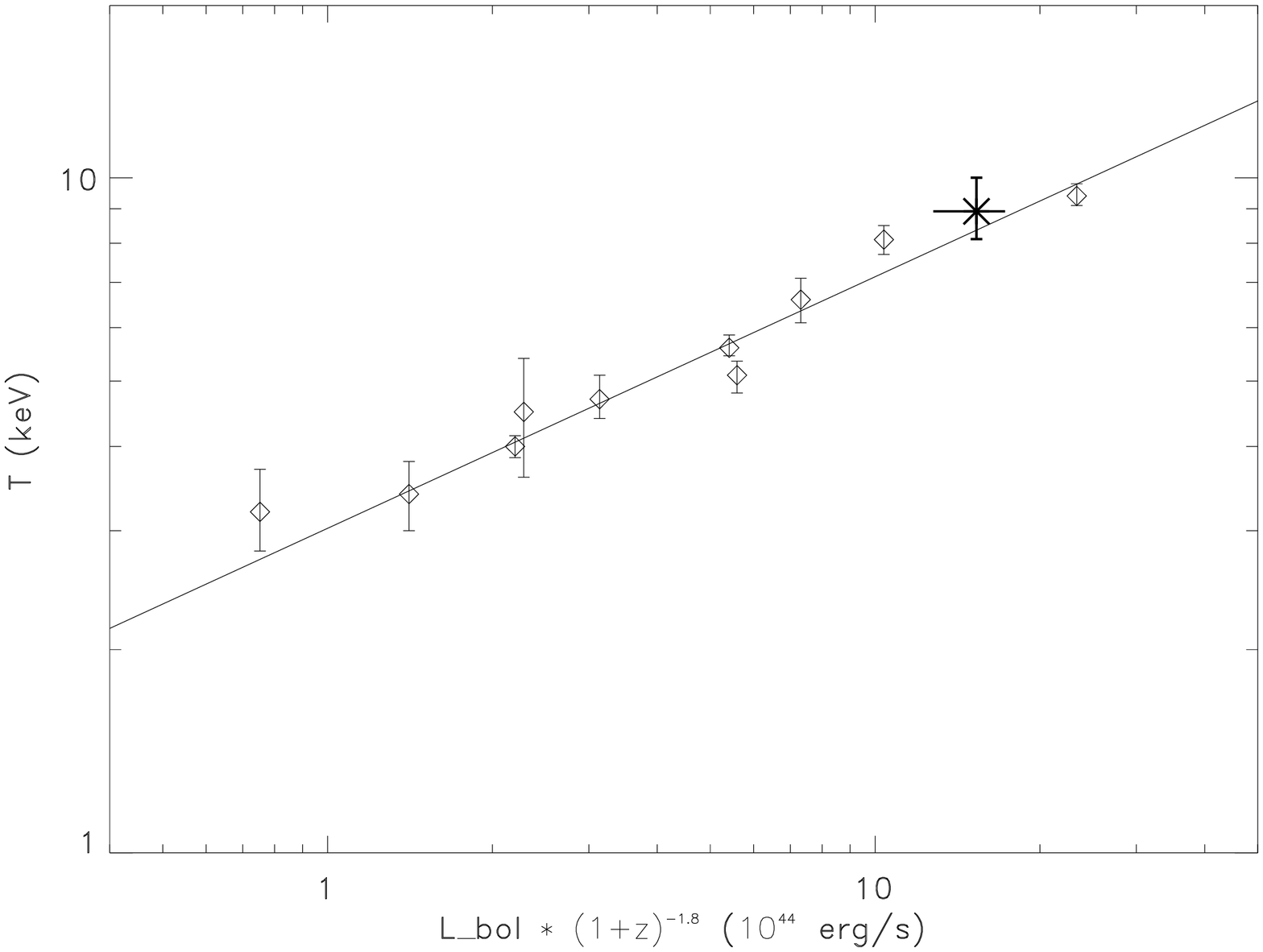,width=6.5cm,angle=90}}
\caption{Comparison of the present result with the $L-T$ relation of Kotov \& 
Vikhlinin (2005). The star indicates the result of RXCJ2228.6+2036, and the 
diamonds and the best-fitted $L-T$ relation line (the solid line) come from 
Kotov \& Vikhlinin (2005). The error bars of the star are at the 68\% 
confidence level.}
\end{figure}

\subsection{$M-Y$ relation}
The integrated SZ flux $Y_{SZ} \propto \int{k_BTn_edV} \propto M_{gas}T$, 
and thus the simplest X-ray analog is defined as $Y_{X}=M_{gas}T$. Kravtsov 
et al. (2006) show that $Y_{X}$ is the best mass proxy with a remarkably low 
scatter and the $M-Y_{X}$ relation is close to the self-similar prediction. 

For RXCJ2228.6+2036, $M_{500}=(1.19\pm0.35)\times10^{15}$ M$_{\odot}$ and
$Y_{X,500}=21.84^{+2.70}_{-2.04}\times10^{14}$ M$_{\odot}$ keV. We plot 
$M_{500}$ versus $Y_X$ in Fig. 14 (shown as a star) and compare it with the 
$M-Y_X$ relations of Zhang et al. (2008) (the solid line), Kravtsov et al. 
(2006) (the dash-dotted line), Nagai et al. (2007) (the dashed line) and Arnaud 
et al. (2007) (the dotted line), which shows a good consistency.

\begin{figure}[ht]
\centerline{\psfig{file=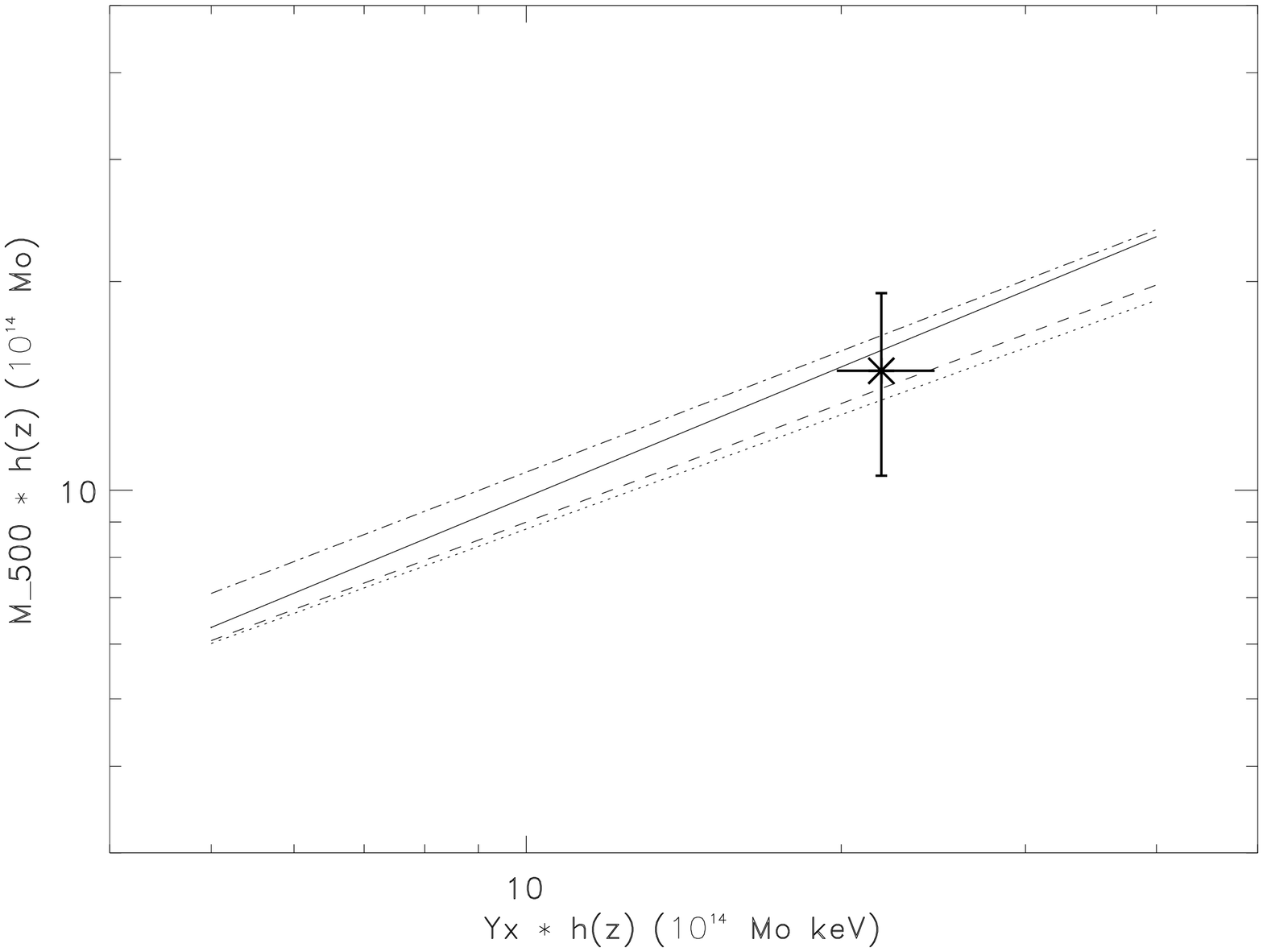,width=6.5cm,angle=90}}
\caption{Comparison of the present result with the $M-Y$ relation of Zhang et 
al. (2008) (the solid line), Kravtsov et al. (2006) (the dash-dotted line), 
Nagai et al. (2007) (the dashed line) and Arnaud et al. (2007) (the dotted line). 
The star indicates the result of RXCJ2228.6+2036 with the errors of 68\% 
confidence level. $h(z)=[0.3(1+z)^3+0.7]^{\frac{1}{2}}$.}
\end{figure}

%%%%%%%%%%%%%%%%%%%%%%%%%%%%%%%%%%%%%%%%%%%%%%%%%%%%%%%%%%%%%%%%%%%%%%%%%%%%%%%
\section{Conclusion}
We presented a detailed analysis of the {\it XMM-Newton} observations of the
distant galaxy cluster RXCJ2228.6+2036 ($z=0.421$) using our deprojection
technique. Through the spectral fitting we derived the deprojected temperature
profile $T(r)$. Weighted by normalizations, we derived a mean temperature
within $r_{500}$, $T_{500}=8.92^{+1.78}_{-1.32}$ keV, which confirms within the 
error bars the previous results by Pointecouteau et al. (2002) and LaRoque 
et al. (2006).
 
Then we calculated the cooling time of this cluster and obtained a cooling
radius of $147\pm10$ kpc. Fitted by a cooling flow model with an isothermal
Mekal component, we derived the mass deposition rate $\sim$
$12.0^{+56.0}_{-12.0}$ M$_{\odot}$yr$^{-1}$ within $r_{cool}$. 

Using the radial density profile $n_e(r)$ and radial temperature profile
$T(r)$, we obtained the mass distribution of RXCJ2228.6+2036. At
$r_{500}=1.61\pm0.16$ Mpc, the total mass is
$M_{500}=(1.19\pm0.35)\times10^{15}$ M$_{\odot}$, in agreement with the
results of Pointecouteau et al. (2002), derived from a combined SZ/X-ray spatial
analysis, and the gas mass fraction is $f_{gas}=0.165\pm0.045$.

We discussed the PSF-correction effect on the spectral analysis and found that
the PSF-corrected temperatures are consistent with those without PSF correction. 

We found a remarkable agreement within the error bars between our X-ray results 
and the SZ measurements in Pointecouteau et al. (2002), which is of prime 
importance for the future SZ survey. The X-ray total mass and X-ray observables 
for RXCJ2228.6+2036 closely obey the empirical scaling relations found in 
general massive galaxy clusters, e.g. the $S$--$T$, $M$--$T$, $L$--$T$ and 
$M$--$Y$ relations, after accounting for self-similar evolution.

%%%%%%%%%%%%%%%%%%%%%%%%%%%%%%%%%%%%%%%%%%%%%%%%%%%%%%%%%%%%%%%%%%%%%%%%%%%%%%%
\begin{acknowledgements}
We thank G. Pratt for providing useful suggestions.
This work was supported by CAS-MPG exchange program. HB and EP acknowledge
support by the DFG for the Excellence Cluster Universe, EXC 153. 
EP acknowledges the support of grant ANR-06-JCJC-0141. YYZ acknowledges support 
from the German BMBF through the Verbundforschung under grant 
No.\,50\,OR\,0601.
\end{acknowledgements}

%%%%%%%%%%%%%%%%%%%%%%%%%%%%%%%%%%%%%%%%%%%%%%%%%%%%%%%%%%%%%%%%%%%%%%%%%%%%%%%

\end{document}